\documentclass[aps,physrev,twocolumn,groupedaddress,superscriptaddress,longbibliography]{revtex4-2}

\usepackage{graphicx}   %
\usepackage{amsmath}
\usepackage{amssymb} 
\usepackage{bm}
\usepackage{mathtools}
\usepackage[braket, qm]{qcircuit} 

\usepackage{xcolor}
\definecolor{apsblue}{rgb}{0.10,0.30,0.65}  %
\usepackage{hyperref}
\hypersetup{
  colorlinks=true,    %
  linkcolor=apsblue,  %
  citecolor=apsblue,  %
  urlcolor=apsblue,   %
  breaklinks=true,    %
} 
  
\usepackage{housestyle}

\begin{document}

\title{Quantum Simulation of the Static Yukawa Model via Effective Particles on a Trapped-Ion Quantum Simulator}

\author{Qingfeng Wang}
\affiliation{Department of Physics and Astronomy, Tufts University, Medford, MA 02155, USA}
\affiliation{Department of Physics, University of Toronto, Toronto, ON M5S 1A7, Canada}

\author{Christopher G. Yale}
\affiliation{Sandia National Laboratories, Albuquerque, NM 87185, USA}

\author{Edward C. Tortorici}
\affiliation{Sandia National Laboratories, Albuquerque, NM 87185, USA}

\author{Brian K. McFarland}
\affiliation{Sandia National Laboratories, Albuquerque, NM 87185, USA}

\author{Daniel Lobser}
\affiliation{Sandia National Laboratories, Albuquerque, NM 87185, USA}

\author{Melissa C. Revelle}
\affiliation{Sandia National Laboratories, Albuquerque, NM 87185, USA}

\author{Andrew J. Landahl}
\affiliation{Sandia National Laboratories, Albuquerque, NM 87185, USA}
\affiliation{Department of Physics and Astronomy, University of New Mexico, Albuquerque, NM 87131, USA}

\author{Susan M. Clark}
\affiliation{Sandia National Laboratories, Albuquerque, NM 87185, USA}

\author{Peter J. Love}
\affiliation{Department of Physics and Astronomy, Tufts University, Medford, MA 02155, USA}
\affiliation{Department of Computer Science, University of Toronto, Toronto, ON M5S 3H6, Canada}
\affiliation{Department of Physics, University of Toronto, Toronto, ON M5S 1A7, Canada}
\affiliation{QMatter, Inc., Office 109, 254 Chapman Rd, Suite 101-B, Newark, DE 19702, USA}
\affiliation{Computational Science Initiative, Brookhaven National Laboratory, Upton, NY 11973, USA}
\begin{abstract}

Trapped-ion, neutral-atom, and circuit QED quantum computers naturally enable coupled qubit-boson simulations.
Boson-fermion interactions arise in quantum field theory, where renormalization is required to obtain finite results. 
We study one of the simplest examples of a coupled fermion-boson system, the static Yukawa model, on the QSCOUT trapped-ion system. 
We use similarity renormalization to express the same physics in terms of effective particles described by different values of the renormalization parameter $s$. 
This enables us to map data taken in the window of bosonic occupancy that is experimentally accessible to larger occupancies of distinct effective particles. 
We use this to study strongly coupled, high-occupancy regimes despite limited readout ranges. 
Using a trapped-ion qubit-boson device on Sandia National Laboratories' QSCOUT platform, we demonstrate the time evolution of the static Yukawa model.
\end{abstract}  
\maketitle

\clearpage

\section{Introduction}\label{sec:intro}

Quantum systems of physical interest, such as quantum field theories, are often intractable for classical computation \cite{bauer2023quantum, bauer2023fundamental}. A quantum simulator makes their simulation efficient by using one controllable quantum system to emulate the dynamics of another \cite{Feynman1982, Lloyd1996}.

Many physical systems involve complex interactions between bosonic and other degrees of freedom, including quantum field theories and lattice gauge theories such as the Schwinger model \cite{schwinger1962, kogutsusskind1975}. Simulating these coupled dynamics has motivated the use of hybrid qubit-boson architectures, which natively pair two-level systems with continuous modes, as realized in circuit QED, trapped-ion, and neutral-atom platforms \cite{blais2021cqed, leibfried2003quantum, katz2023programmable, kaufman2012, shaw2025}.

A substantial body of work has targeted the simulation of fermion-boson field theories, in which the fermionic and bosonic degrees of freedom map directly onto the qubits and their associated modes. It spans theoretical frameworks \cite{casanova2011qft, crane2024, liu2026} and experimental realizations \cite{zhang2018fermion}. The closest to the present work is the recent observation of real-time (1+1)-dimensional Yukawa dynamics \cite{than2026observation}.

However, characterizing such systems at high occupation is limited by the readout rather than the simulation. A qubit-boson platform hosts high-occupation states natively, without the qubit overhead that a register encoding would demand \cite{klco2019digitization}, but the range over which the occupation can be resolved remains bounded. Although these platforms can prepare occupations of order a hundred quanta \cite{mccormick2019, deng2023largefock}, number-resolved readout of an unknown occupation has been demonstrated only up to occupations of order ten quanta \cite{than2026observation, schuster2007resolving, brown2022tomography}.
 
A concrete example is the static Yukawa model, the simplest coupling of a fermion to a single bosonic mode. Unlike Yukawa theory, which requires renormalization \cite{serafin2025secondorder, gustin2025renormalized}, the static model is finite and exactly solvable \cite{glazek2021elementary}. This exact solvability provides a controlled benchmark, and as the coupling grows the bosonic occupation rises beyond the resolvable range.

To bring these high occupations within the readout range, we recast the dynamics in terms of effective particles. We apply the renormalization group procedure for effective particles (RGPEP), a form of similarity renormalization that parameterizes the system by a flow parameter $s$~\cite{glazek2021elementary, glazek2020abelian, glazekmaslowski2002poincare}. This transformation preserves the energy spectrum while systematically lowering the occupation of the effective particles. Consequently, a measurement performed within the reliable readout range at nonzero $s$ maps, through a known transformation, back to the high-occupation regime of the original bare particles.

In this work, we realize static Yukawa Hamiltonians under RGPEP flow on the QSCOUT trapped-ion quantum computer \cite{clark2021qscout}. We reproduce the time evolution where the occupation is reliably resolved and locate where the readout breaks down as strong coupling drives the occupation beyond that range. Raising the flow parameter $s$ then maps this strongly coupled regime onto effective particles of lower, measurable occupation, whose values follow the known flow and render the regime accessible through in-range measurements.

The remainder of this paper is organized as follows.
\cref{sec:theory} develops the RGPEP description of the static Yukawa model and shows how the occupation scales with the flow parameter $s$.
\cref{sec:method} describes how the time evolution compiles into a circuit on QSCOUT and how the occupation is read out from the sideband oscillations of a probe qubit.
\cref{sec:results} presents the measured occupation as we scan the evolution time, the coupling, and the flow parameter.
Finally, \cref{sec:discussion} interprets these results, including how an in-range measurement recovers the out-of-range regime, and open challenges beyond this benchmark.  

\section{Theory}\label{sec:theory}
In this section we explain RGPEP theory as applied to the static Yukawa model.

\subsection{The static Yukawa model}
\label{sec:static_yukawa_concept}

Yukawa theory has recently been simulated on spatial lattices using digital Trotterization \cite{than2026observation}. 
Here we focus on a version of this theory known as the static Yukawa model \cite{glazek2021elementary}.

 The Hamiltonian $H$ is
\begin{align}
  \label{eqn:static-yukawa}
  H = E_{\mathrm{f}} b^\dagger b + E_{\mathrm{b}} a^\dagger a + gE_\mathrm{I} b^\dagger b (a^\dagger + a),
\end{align}
where $E_{\mathrm{f}}$, $E_{\mathrm{b}}$, and $E_\mathrm{I}$ are the fermion, boson, and interaction energy parameters, respectively \cite{glazek2021elementary}. The dimensionless coupling constant is $g$, and the operators $b^{(\dagger)}$ and $a^{(\dagger)}$ annihilate (create) the static fermion and boson.

The bosons obey the standard commutation rules:
\begin{align}
  [a, a^\dagger] = 1; \quad [a,a]=[a^\dagger,a^\dagger]=0,
\end{align}
while the fermions obey the standard anti-commutation rules \cite{glazek2021elementary}:
\begin{align}
  \{b, b^\dagger\} = 1; \quad \{b,b\}=\{b^\dagger,b^\dagger\}=0,
\end{align}
and any combination of bosonic and fermionic operators commutes:
\begin{align}
  [a^{(\dagger)}, b^{(\dagger)}] = 0.
\end{align}
Next we define the RGPEP procedure for this model.

\subsection{RGPEP formalism}\label{sec:model}

We define the original Hamiltonian $H$, \cref{eqn:static-yukawa}, as the bare Hamiltonian which is constructed from particle operators
\begin{align}
   q \in \{a,a^\dagger, b, b^\dagger\}.
\end{align}

The goal of the renormalization group method is to define
\begin{align}
  H = H_s
\end{align}
such that the Hamiltonian is represented by effective particle operators  $q_s$ parametrized by a flow parameter $s$.
The basis transformation $\mathcal{U}_s$ converts between the bare and effective particle operators via
\begin{align}
  q_s = \mathcal{U}_s q \mathcal{U}_s ^\dagger .   
\end{align}

In the RGPEP formalism~\cite{glazekmaslowski2002poincare, glazek2020abelian,glazek2021elementary}, $\mathcal{H}_s$ is the solution of
\begin{align}
  \frac{d}{ds} \mathcal{H}_s &= [\mathcal{G}_s,\mathcal{H}_s] \label{eqn:flow}\\
  \mathcal{G}_s &:= [H_{\mathrm{f}}+H_{\mathrm{b}}, \mathcal{H}_s] \label{eqn:generator}
\end{align}
with initial condition $\mathcal{H}_0= H$, where
$H_{\mathrm{b}}$ and $H_{\mathrm{f}}$ are the free-boson and free-fermion Hamiltonians, respectively.
$\mathcal{H}_s$ is the renormalized Hamiltonian such that 
\begin{align} 
  \label{eqn:Hs-Us}
   \mathcal{H}_s  = \mathcal{U}_s^\dagger H_s   \mathcal{U}_s.
\end{align}
The operator $\mathcal{U}_s$ is recovered from the generator $\mathcal{G}$ in the integral form
\begin{align}
  \label{eqn:Us}
  \mathcal{U}_s^\dagger =  \mathcal{T}  \exp \left( \int_{0}^{s} d\tau\, \mathcal{G}_\tau \right),
\end{align}
where $\mathcal{T}$ represents the ordering in $\tau$.
At $s = \infty$, the effective particles become the physical particles of the model.

\subsection{RGPEP applied to the static Yukawa model}\label{sec:yukawa}

In the RGPEP formalism, we extract the free component from the static Yukawa Hamiltonian \cref{eqn:static-yukawa}, 
\begin{align}
    H_{\mathrm{f}} = E_{\mathrm{f}} b^\dagger b; \quad
  H_{\mathrm{b}} = E_{\mathrm{b}} a^\dagger a.
\end{align}

The RGPEP \cref{eqn:flow,eqn:generator,eqn:Us} can be solved exactly for the static Yukawa model~\cite{glazek2021elementary}, giving
\begin{align}
    \mathcal{U}_s^\dagger&=e^{ c_s b^\dagger (a^\dagger -a )b};\\
  a_s = a + &c_s b^\dagger b; \quad b_s = e^{c_s (a^\dagger -a)}b\\
  a_s^\dagger = a^\dagger + &c_s b^\dagger b; \quad b_s^\dagger = e^{-c_s (a^\dagger- a)}b^\dagger
\end{align}
where $s$-dependent parameters are
\begin{align}
  c_s &=  (g - g_s) E_\mathrm{I}/E_{\mathrm{b}},\\
  g_s &= g e^{-E_{\mathrm{b}}^2 s}. \label{eqn:gs}
\end{align}
See \cite{glazek2021elementary} for details of the derivation.

\subsection{Hamiltonians in different bases}

\begin{figure}[t]
  \centering
  \includegraphics[width=\columnwidth]{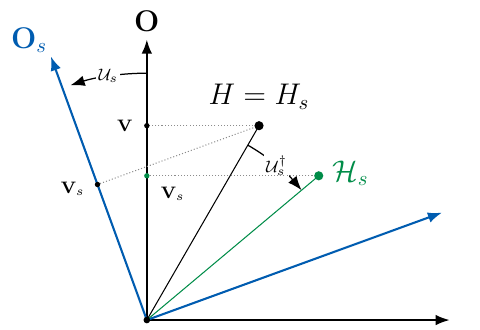}
  \caption{Vector-space relationship among the bare ($H$), effective ($H_s$), and renormalized ($\mathcal{H}_s$) Hamiltonians in a two-basis example.
RGPEP defines $H=H_s = \mathcal{U}_s \mathcal{H}_s \mathcal{U}_s^\dagger$.
The bare basis axis $\mathbf{O}$ is shown in black and the effective basis axis $\mathbf{O}_s$ in blue, 
connected by the unitary rotation $\mathcal{U}_s$.
The components of $H$ in the bare and effective bases are $\mathbf{V}$ and $\mathbf{V}_s$, respectively.
$\mathcal{H}_s$ is instead obtained by the rotation $\mathcal{U}_s^\dagger$.  
Specifically, for the static Yukawa model, $\mathcal{H}_s = \mathbf{V}_s \cdot \mathbf{O}$, shown in green.  }  
  \label{fig:static-yukawa}
\end{figure}

To better compare the Hamiltonians,
we write them in the form
\begin{align}
H = \mathbf{V}\cdot \mathbf{O}
\end{align}
where $\mathbf{V}$ is a vector of scalar coefficients and
$\mathbf{O}$ is a vector of operators.  
The bare Hamiltonian $H$ is then defined via
\begin{align}
\begin{array}{r @{{}={}} c c c l}
  \mathbf{V}          & [\, E_{\mathrm{f}}         & E_{\mathrm{b}}         & g E_\mathrm{I}                        & ];\\
\mathbf{O} & [\, b^\dagger b & a^\dagger a & (a^\dagger + a)\, b^\dagger b & ]^T.
\end{array}
\end{align}
With the unitary rotation
\begin{align}
 \mathbf{O}_s  = \mathcal{U}_s \mathbf{O}   \mathcal{U}_s^\dagger,
\end{align}
the effective Hamiltonian $H_s$ is then
\begin{align}
  H_s = \mathbf{V}_s \cdot \mathbf{O}_s
\end{align}
 with
\begin{align}
\begin{array}{r @{{}={}} l c c c l}
  \mathbf{V}_s          & [& E_{\mathrm{f}} + g_s^2 \Delta_s          & E_{\mathrm{b}}         & g_s E_\mathrm{I}                        & ];\\
\mathbf{O}_s & [& b^\dagger_s b_s & a^\dagger_s a_s & (a^\dagger_s + a_s)\, b^\dagger_s b_s & ]^T,
\end{array}
\end{align}
where $\Delta_s = \left(1 - e^{2E_{\mathrm{b}}^2 s} \right) E_\mathrm{I}^2/E_{\mathrm{b}}$.
Note the coefficient vector $\mathbf{V}_s$ is also parametrized by $s$ and modified accordingly after the basis transformation to satisfy the constraint $H_s = H$.

When $s = \infty$, $H_s$  by definition becomes the physical Hamiltonian, 
\begin{align}
  H_\infty = \mathbf{V}_\infty \cdot \mathbf{O}_\infty.
\end{align}
For a general model, $H_\infty$ has a band-diagonal structure.
However, in the static Yukawa model, the fermion-boson coupling $g_s$ is completely suppressed according to \cref{eqn:gs}, and
\begin{align}
\begin{array}{r @{{}={}} l  c c l}
  \mathbf{V}_\infty          & [& E_{\mathrm{f}} - g^2 E_\mathrm{I}^2/E_{\mathrm{b}}          & E_{\mathrm{b}}                               & ];\\
\mathbf{O}_\infty & [& b^\dagger_s b_s & a^\dagger_s a_s   & ]^T,
\end{array}
\end{align}
which is simply a diagonal form, and the eigenenergies can be extracted directly from the coefficients.

Finally, the renormalized Hamiltonian $ \mathcal{H}_s$ takes a different operator basis $\mathbf{\bar{O}}_s $ rotated in reverse order, 
\begin{align}
\mathbf{\bar{O}}_s  = \mathcal{U}_s^\dagger \mathbf{O}   \mathcal{U}_s,
\end{align}
and therefore
\begin{align}
 \mathcal{H}_s = \mathbf{V}\cdot \mathbf{\bar{O}}_s.
\end{align}
In general $\mathbf{\bar{O}}_s$ has a nontrivial relationship to $\mathbf{O}_s$, and to
time-evolve $ \mathcal{H}_s$ we have to implement the unitary transformation
\begin{align}
  e^{-i \mathcal{H}_s t} = \mathcal{U}_s^\dagger e^{-iH_s t} \mathcal{U}_s.
\end{align}
However, in the static Yukawa model, the flow only displaces the boson, conditioned on the fermion number, which keeps the operator basis closed~\cite{glazek2021elementary} and gives
\begin{align} 
  \label{eqn:mathcal_H}
  \mathcal{H}_s = \mathbf{V}_s \cdot \mathbf{O},
\end{align}
 so time-evolving under $\mathcal{H}_s$ is simplified to time-evolving under $H$ with different coefficients.

The relationship between the Hamiltonians and their composite operator basis is illustrated in \cref{fig:static-yukawa}.

\subsection{Ground state of the static Yukawa model} 
\label{subsec:ground-state}
We focus on the sector where the number of fermions is exactly one.  
The ground state of the static Yukawa model~\cite{glazek2021elementary}, $|\mathrm{G}\rangle$, in the bare basis is then
\begin{align}
  |\mathrm{G}\rangle = |1\rangle_{\mathrm{f}} \otimes |\alpha\rangle_{\mathrm{b}},
\end{align}
where subscripts $\mathrm{f}$ and $\mathrm{b}$ stand for the fermionic and bosonic components, respectively, and $|\alpha\rangle_{\mathrm{b}}$ 
is a coherent state with $\alpha = -g  E_\mathrm{I}/E_{\mathrm{b}}$.

The ground state can be determined entirely by the operators applied to the vacuum state
\begin{align}
  |\mathrm{vac}\rangle =  |0\rangle_{\mathrm{f}}\otimes |0\rangle_{\mathrm{b}}.
\end{align}
As a result, we can also write
\begin{align}
  \label{eqn:gs-state}
  |\mathrm{G}\rangle = D(\alpha)  b^\dagger |\mathrm{vac}\rangle
\end{align}
where $D(\alpha)$ is a  displacement operator 
\begin{align}
   D(\alpha)  =  e^{\alpha a^\dagger - \alpha^* a}.
\end{align}
This enables the generalization of the ground state to an effective version at scale $s$ 
\begin{align}
  |\mathrm{G}\rangle_s  =D(\alpha_s)  b^\dagger_s    |\mathrm{vac}\rangle,
\end{align}
where 
\begin{align}
  \alpha_s = \alpha_s^* = -g_s  \frac{E_\mathrm{I}}{E_{\mathrm{b}}}
\end{align}
is real-valued.

When $s=\infty$, the physical ground state is then
\begin{align}
  |\mathrm{G}\rangle_\infty  = b^\dagger_\infty  |\mathrm{vac}\rangle
\end{align}
since $ D(\alpha_\infty)  = I$. 
In this case, the ground state is fully determined by a single physical fermion 
\begin{align}
  b^\dagger_\infty =D(\alpha) b^\dagger
\end{align}
which is mathematically equivalent to \cref{eqn:gs-state}.
However, conceptually $|\mathrm{G}\rangle_\infty $ contains only a single physical fermion, which, expressed in the bare basis, is a bare fermion accompanied by a coherent cloud of bare bosons. This $s\to\infty$ physical particle coincides with the ``clothed'' particle of Greenberg and Schweber~\cite{greenbergschweber1958clothed}, reached here through the continuous flow in $s$ rather than a single transformation.

\subsection{Time evolution at different flow parameters}
We study the time evolution of the system with the bare occupation number operator 
\begin{align}
  N\equiv a^\dagger a,
\end{align}
 not the effective number $a^\dagger_s a_s$.
Throughout, a subscript $H$ or $\mathcal{H}$ on the expectation $\langle\,\cdot\,\rangle$ labels the evolution frame as the bare or renormalized Hamiltonian, respectively. The bare occupation is a time-dependent scalar and has the form
\begin{align}
  \langle N\rangle_H(s,t)
  =\langle\mathrm{init}|\,e^{iHt}\, (a^\dagger a)\,e^{-iHt}\,|\mathrm{init}\rangle
\end{align}
where $|\mathrm{init}\rangle$ is the initial state. 
Similarly, the bare occupation for the renormalized Hamiltonian $\mathcal{H}_s$ is
\begin{align}
 \langle N\rangle_{\mathcal{H}}(s,t) =
  \langle \mathrm{init}|\,e^{i\mathcal{H}_s t}\,  (a^\dagger a)\,e^{-i\mathcal{H}_s t}\,|\mathrm{init}\rangle,
  \label{eqn:Ns-meas}
\end{align}
which is also a time-dependent scalar, obtained by updating only the time evolution.
The renormalized-frame $\langle N\rangle_{\mathcal{H}}$ is what we actually measure on the quantum computer.

In general, there is a nontrivial map from $\langle N\rangle_{\mathcal{H}}$ to $\langle N\rangle_H$ for recovering the dynamics,
\begin{align}
\langle N\rangle_H = \langle N\rangle_{\mathcal{H}}& + 2\,\mathrm{Re}\big\langle\mathrm{init}|\, e^{i\mathcal{H}_s t} a^\dagger e^{-i\mathcal{H}_s t} \Xi_s \,|\mathrm{init}\rangle \notag \\
    &+ \langle\mathrm{init}|\, \Xi_s^\dagger \Xi_s  \,|\mathrm{init}\rangle
\end{align}
where 
\begin{align}
\Xi_s(t) = e^{iHt} a\, e^{-iHt} - e^{i\mathcal{H}_s t} a\, e^{-i\mathcal{H}_s t}.
\end{align}
See \cref{app:general} for the derivation of this expression.
However, for the static Yukawa model this map can be computed analytically (see \cref{eqn:recover} for the explicit recovery in practice).

\section{Method}\label{sec:method}
Here we explain the end-to-end workflow from time-evolution decomposition to post-processing, organized into two subsections.
\cref{subsec:circuit-decomp} explains how the time-evolution unitary is decomposed into an executable circuit tailored to the QSCOUT native gate set.
In \cref{sec:experiment} we show how to read out the occupation $\langle N\rangle_{\mathcal{H}}$ and obtain the error bar.

On the QSCOUT platform, the boson is realized as a shared motional mode of the trapped-ion chain, while the qubit operations act on the ions' hyperfine ``spins.'' These hyperfine qubits are used as ancillae, one that synthesizes the displacement and a separate probe for readout, and are distinct from the model's static fermion of \cref{sec:theory}.

\subsection{Circuit decomposition}
\label{subsec:circuit-decomp}

 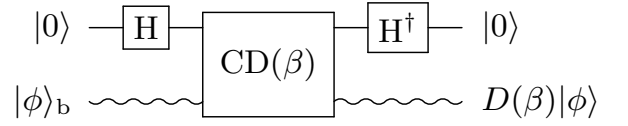
\begin{figure}[t]  
    \centering
    \hspace*{-6mm}
    \resizebox{0.3\textwidth}{!}{%
    \Qcircuit @C=1em @R=1em {
\lstick{|0\rangle}     &\gate{\mathrm{H}}       &\multigate{1}{{\mathrm{CD}}(\beta)}    &\gate{\mathrm{H}^\dagger}       &\qw   &\rstick{\hspace*{-3mm}|0\rangle}            \\
\lstick{|\phi\rangle_{\mathrm{b}}}       &                               & \ghostc{{\mathrm{CD}}(\beta)}[-2]                        &                           & \qwc[-2]  &\rstick{\hspace*{-3mm}D(\beta)|\phi\rangle}         }
    }
    \caption{Circuit compilation for displacement operator $D(\beta)$ realized via a conditional-displacement operator  $\mathrm{CD}(\beta)$.
    The top straight line represents the qubit initialized with $|0\rangle$ and the bottom wavy line represents the boson, prepared in an arbitrary single-mode state $|\phi\rangle_{\mathrm{b}}$.
    The net effect is the application of  $D(\beta)$ in the bosonic mode. In practice, to mitigate bit-flip errors, we retain only the shots in which the qubit is measured in $|0\rangle$.
    }
        \label{fig:circuit}
  \end{figure}

\begin{figure*}[ht]
  \centering
  \includegraphics[width=\textwidth]{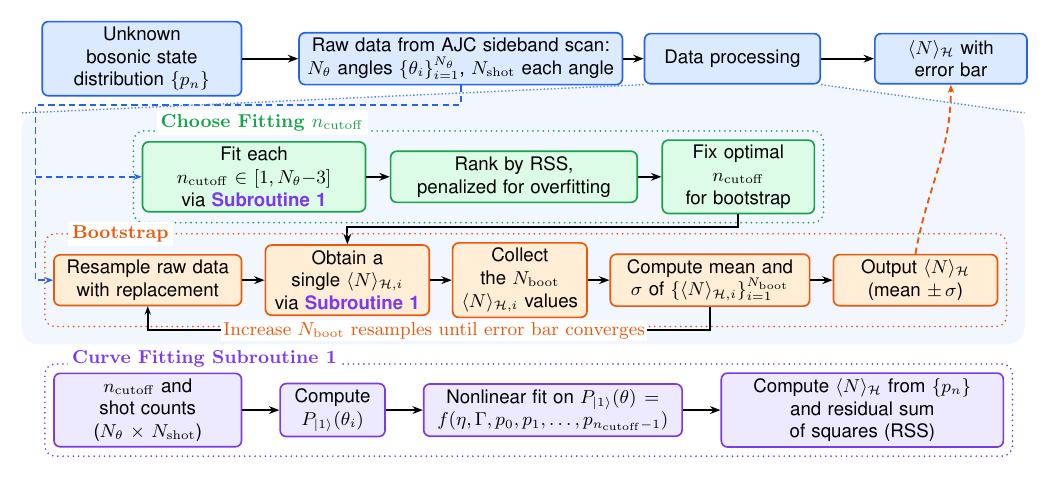}
  \caption{End-to-end readout workflow that turns one anti-Jaynes-Cummings (AJC)
  sideband scan into a single occupation $\langle N\rangle_{\mathcal{H}}$ with an
  error bar. The top row (blue) is the high-level pipeline: an unknown Fock
  distribution $\{p_n\}$ is probed by an AJC scan over $N_\theta$ angles with
  $N_{\rm shot}$ shots each, and the recorded shots are processed into
  $\langle N\rangle_{\mathcal{H}}$. The shaded region below, linked to the
  ``Data processing'' box by the dotted lines, expands that box into the two stages that
  produce $\langle N\rangle_{\mathcal{H}}$, both of which call the shared curve-fitting routine drawn in
  purple (\emph{Subroutine 1}). Given a cutoff $n_{\rm cutoff}$ and the shot counts,
  this routine computes the probabilities $P_{|1\rangle}(\theta_i)$,
  performs the non-linear sideband fit
  $f(\eta,\Gamma,p_0,\dots,p_{n_{\rm cutoff}-1})$ of \cref{eq:sideband}, and
  returns a single $\langle N\rangle_{\mathcal{H}}$ with the residual sum of squares (RSS).
  The green stage selects the fit cutoff: it runs the routine for every
  $n_{\rm cutoff}\in[1,N_\theta-3]$, ranks the fits by RSS penalized by the remaining
  degrees of freedom (reduced $\chi^2$, \cref{app:readout}), and fixes the optimal
  $n_{\rm cutoff}$. 
  The orange bootstrap stage builds the error bar: holding that
  $n_{\rm cutoff}$ fixed, it resamples the recorded shots with replacement and
  refits each resample through the same \emph{Subroutine 1}, accumulating $N_{\mathrm{boot}}$ values of
  $\langle N\rangle_{\mathcal{H}}$ and increasing $N_{\mathrm{boot}}$ until their mean and standard
  deviation converge (\cref{app:error}). The reported point is the bootstrap mean
  and the error bar is one standard deviation, returned to the top row along the
  dashed orange arrow. Solid arrows trace the processing sequence,
  and dashed blue arrows carry the raw scan data into both stages.}
  \label{fig:workflow}
\end{figure*}

To simplify the time evolution of $\mathcal{H}_s$, we restrict to the sector where there is a single fermion. 
According to \cref{eqn:mathcal_H},  time-evolution is expanded as
\begin{align}
  U(t) = e^{-i \left( (E_{\mathrm{f}} + g_s^2 \Delta_s)
+ E_{\mathrm{b}} a^\dagger a + g_s E_\mathrm{I}(a^\dagger + a) \right) t}.
\end{align}
Prior realizations of Yukawa-type fermion-boson dynamics digitize the time evolution with a Trotter-Suzuki expansion~\cite{kaldenbach2024yukawa, than2026observation}, or engineer it directly by analog embedding~\cite{zhang2018fermion}. 
Due to the special structure of the Hamiltonian, we instead compile its renormalized evolution exactly into a three-part decomposition
\begin{align}
  U(t) = P(t)R(t)D(t),
\end{align}
where 
\begin{align}
  P(\phi_t) = e^{-i \phi_t}
\end{align}
is a global phase, 
\begin{align}
  R(\theta_t) = e^{-i \theta_t a^\dagger a}
\end{align}
is a rotation operator and
\begin{align}
 D(\beta_t) = e^{\beta_t a^\dagger - \beta_t^* a}
\end{align}
is the displacement operator. The time-dependent  parameters are 
\begin{align}
  \phi_t &= (E_{\mathrm{f}} + g_s^2 \Delta_s) t - \frac{g_s^2 E_\mathrm{I}^2}{E_{\mathrm{b}}^2} (E_{\mathrm{b}} t - \sin(E_{\mathrm{b}} t))\\
  \theta_t &= E_{\mathrm{b}} t\\
  \beta_t &= \alpha_s \left(e^{i E_{\mathrm{b}} t} - 1\right),
\end{align}
where $\alpha_s = -g_s E_\mathrm{I}/E_{\mathrm{b}}$ is the real static ground-state displacement of \cref{subsec:ground-state}.
This exact decomposition is the Wei-Norman form of the evolution operator~\cite{wei1963lie}, derived in \cref{app:prd}.

Global phase has no observable effect, so we can ignore $P(t)$.
In a full implementation, we would apply $D(t)$ followed by the $R(t)$ operation.
However, because the final $R(t)$ commutes with the occupation number operator, it has no observable effect on the measured occupation and can also be safely ignored.
As a result, the entire circuit simplifies to a single $D(\beta_t)$.

QSCOUT's trapped-ion quantum computer currently does not support pure boson gates such as $D(\cdot)$ directly. 
However, it features a conditional displacement (CD) gate, realized via a spin-dependent force~\cite{sorensen1999quantum, sorensen2000entanglement}
\begin{align}
\mathrm{CD}(\beta)  &=e^{X(\beta a^\dagger -\beta^* a)},
\end{align}
where $X$ is the Pauli-$X$ operator of the ancilla qubit.
As a result, $D(\beta)$ can be compiled from $\mathrm{CD}(\beta)$ with an ancilla qubit and a basis transformation.
See \cref{fig:circuit} for the circuit diagram.
The Hadamard gate can be further decomposed based on the following identity
\begin{align}
  \label{eqn:h-decom}
  \mathrm{H} =e^{i \frac{\pi}{2}} R_z \left(\frac{\pi}{2} \right) R_x \left(\frac{\pi}{2} \right) R_z \left(\frac{\pi}{2} \right).
\end{align}

\subsection{Occupation readout}\label{sec:experiment}

In each experimental run we time evolve under the static Yukawa Hamiltonian for time $t$ and obtain the average occupation
$\langle N\rangle_{\mathcal{H}}$ from the
distribution $\{p_n\}$ of each Fock level $n$ using
a single probe qubit via an anti-Jaynes-Cummings (AJC), or blue-sideband, interaction~\cite{jaynes1963comparison, leibfried2003quantum}:
\begin{align}
  \mathrm{AJC}(\theta,\phi) = \exp\big[-i\frac{\theta}{2}(e^{-i\phi}\sigma_+a^\dagger+e^{i\phi}\sigma_-a)\big].
\end{align}
 \cref{fig:workflow} summarizes the readout and data-processing procedure.

AJC simultaneously drives all isolated two-level subspaces to oscillate at distinct frequencies,
\begin{align}
  \label{eq:oscillate}
    |0\rangle\otimes |n\rangle_{\mathrm{b}}  \xleftrightarrow[\text{Frequency}\, \propto \sqrt{n+1}]{\text{ Driven by AJC}} |1\rangle\otimes |n+1\rangle_{\mathrm{b}},
\end{align}
giving an analytical relation between the probe qubit dynamics and $\{p_n\}$,
\begin{align}
  P_{|1\rangle}(\theta)=\frac{1}{2}-\frac{1}{2}\,e^{-\Gamma \theta}\sum_{n=0}^{n_{\rm cutoff}-1}p_n\cos\!\big(\eta\sqrt{n+1}\, \theta \big),
  \label{eq:sideband}
\end{align}
where $P_{|1\rangle}(\theta)$ is the probability of measuring $|1\rangle$ in the probe qubit, $\Gamma \ge 0$ is the decoherence rate, $\eta \in [0.85, 1.05]$ accounts for calibration error and relative ion-beam positional drift, and $n_{\rm cutoff}$ is the number of $\{p_n\}$ parameters to be fitted.
See \cref{app:readout} for derivation.

The readout procedure involves driving the sideband oscillation via AJC$(\theta_i)$ to a specific angle $\theta_i$ and measuring $P_{|1\rangle}(\theta_i)$ using $N_\mathrm{shot}$ sampling shots. 
This process is repeated across $N_\theta$ different angles to collect the dataset $\{P_{|1\rangle}(\theta_i)\}_{i=1}^{N_\theta}$. 
We then perform non-linear fitting to extract $\{p_n\}$ alongside the hardware-related parameters $\eta$ and $\Gamma$. 
The maximum number of quanta in the fit is $n_{\rm cutoff}-1$, which is chosen adaptively based on the fitting quality.  The procedure for choosing $n_{\rm cutoff}$ is detailed in \cref{app:readout}.
 
The fit does not impose $\sum_n p_n = 1$.  
Because the occupation depends only on relative populations, we report the normalized mean
\begin{align}
  \langle N\rangle_{\mathcal{H}} =  \frac{\sum_n n\,p_n}{\sum_n p_n}.
\end{align}

The error bar on each $\langle N\rangle_{\mathcal{H}} $ is obtained by bootstrapping the raw data shots and reported as the mean and standard deviation of the resampled values. 
The resampling construction and the adaptive choice of the number of resamples are detailed in \cref{app:error}.
The obtained error bar represents the statistical uncertainty caused by shot noise, not the quantum computer's intrinsic hardware errors.

\section{Experimental Results}\label{sec:results}
\begin{figure}[t]
  \centering
  \includegraphics[width=\columnwidth]{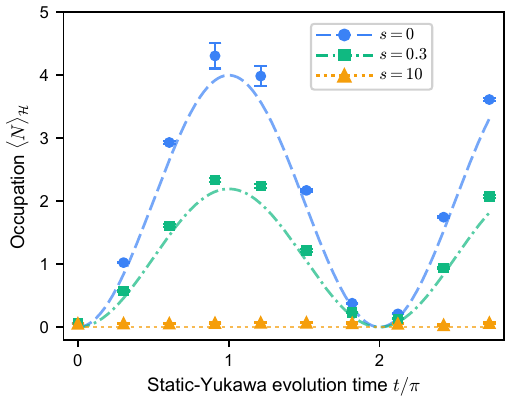}
  \caption{Time scan of the static Yukawa evolution $e^{-i \mathcal{H}_s t}$ at
    fixed coupling $g=1$ and three flow parameters $s=\{0,\,0.3,\,10\}$. Markers are
    experiment ($N_\theta=100$, $N_\mathrm{shot}=4800,\,2400,\,300$,
    respectively) and error bars are the bootstrap standard deviation. Curves are the coherent-state prediction $4g^2 e^{-2s}\sin^2(t/2)$ in \eqref{eqn:boson-number}, each drawn in the line style paired with its markers.
    The initial state has exactly one fermion and zero
    bosons. Increasing $s$ scales the oscillation amplitude down by $e^{-2s}$.}
  \label{fig:dynamics}
\end{figure} 

The initial state is the bosonic vacuum:
\begin{align}
  |\mathrm{init}\rangle =  |1\rangle_{\mathrm{f}}\otimes |0\rangle_{\mathrm{b}}.
\end{align}
In addition, we set $E_{\mathrm{b}}=E_{\mathrm{f}}=E_\mathrm{I} = 1$.
As a result, the time evolution depends only on the evolution time $t$, the coupling strength $g$, and the flow parameter $s$.
More concretely, from the circuit decomposition of \cref{subsec:circuit-decomp}, it reduces to a displacement operator $D(\beta)$ with
\begin{align}
\beta(g,s,t) &= -g_s\,(e^{it}-1),\label{eqn:beta-gs-connection}\\
g_s &= g\,e^{-s} ,
\end{align}
and the occupation is
\begin{align}
  \label{eqn:boson-number}
  \langle N\rangle_{\mathcal{H}}(g,s,t) = 4g^2_s   \sin^2(t/2).
\end{align}
To recover the original bare version,
\begin{align}
  \label{eqn:recover}
  \langle N\rangle_H(g,s, t) = \left(\frac{g}{g_s}\right)^2 \langle N\rangle_{\mathcal{H}}(g,s,t).
\end{align}

The derivation is shown in \cref{app:yukawa}.
We then perform experiments by scanning $t$, $g$ or $s$ while keeping the other two parameters fixed.

\subsection{Scanning time $t$}

We first demonstrate the time evolution of the static Yukawa model at three selected flow parameters $s=\{0,0.3,10\}$
and keep $g=1$, shown in
\cref{fig:dynamics}. At $s=10$ the dynamics are practically those of $s=\infty$, since the current readout cannot resolve the difference between the two.

The three measured curves follow the theory qualitatively (\cref{fig:dynamics}).
This is the expected behavior when the occupation stays within the reliable readout range.

Each occupation curve can be conceptually understood as the time evolution of the same system described by an effective fermion at a different flow scale $s$.
As the flow parameter $s$ increases, more bare bosons are absorbed into the effective fermion, lowering the amplitude.
The $s=10$ dynamics approximate the physical limit, in the sense that the entire system is a single physical fermion with no physical boson and hence zero occupation.

\begin{figure*}[t]
  \centering
  \includegraphics[width=\textwidth]{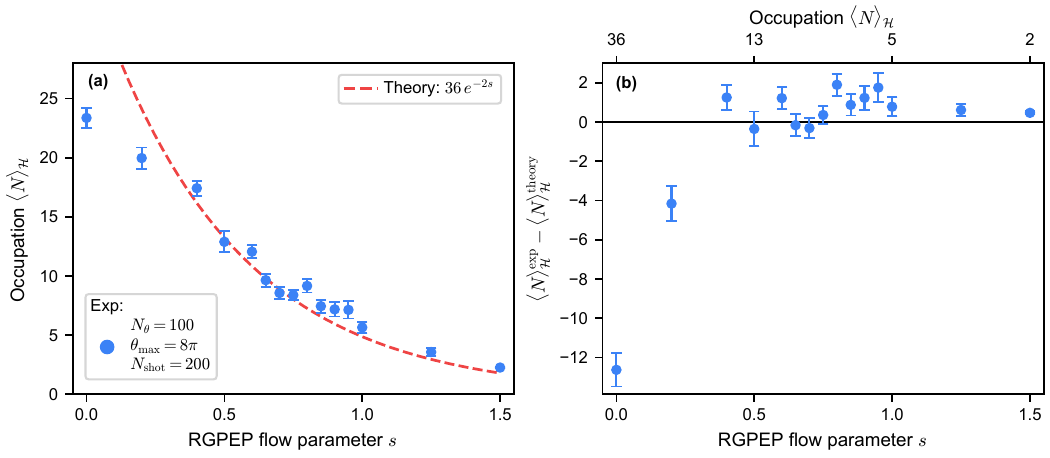}
  \caption{Flow scan of the static Yukawa evolution $e^{-i \mathcal{H}_s t}$ at fixed $t = \pi$ and strong coupling $g=3$. The initial state has exactly one fermion and zero bosons.
  (a) Occupation $\langle N\rangle_{\mathcal{H}}$ versus flow parameter $s$, with the coherent-state prediction $36\,e^{-2s}$ (dashed) given by \eqref{eqn:boson-number}.
  (b) Residual $\langle N\rangle_{\mathcal{H}}^{\rm exp}-\langle N\rangle_{\mathcal{H}}^{\rm theory}$ versus $s$. The top axis reads the corresponding $\langle N\rangle_{\mathcal{H}}=36\,e^{-2s}$.
  Only points with $\langle N\rangle_{\mathcal{H}}\gtrsim 1$ are shown. Error bars are the bootstrap standard deviation. }
  \label{fig:flow} 
\end{figure*}

We also observe a small but consistent positive offset of the measured occupation $\langle N\rangle_{\mathcal{H}}$ from its theoretical prediction within the readable range, visible in the residual panels of the coupling and flow scans (\cref{fig:scan-g}b and \cref{fig:flow}b). 
The offset is systematic rather than statistical.
Because the readout ceiling can only bias the occupation downward, the offset's positive sign rules out that mechanism.
We attribute it to residual hardware imperfections, including an initial residual phonon number, $\bar{n}\sim0.1$, and do not correct for it.
The qualitative agreement with theory is unaffected.

\subsection{Scanning coupling strength $g$}

\begin{figure*}[t]
  \centering 
  \includegraphics[width=\textwidth]{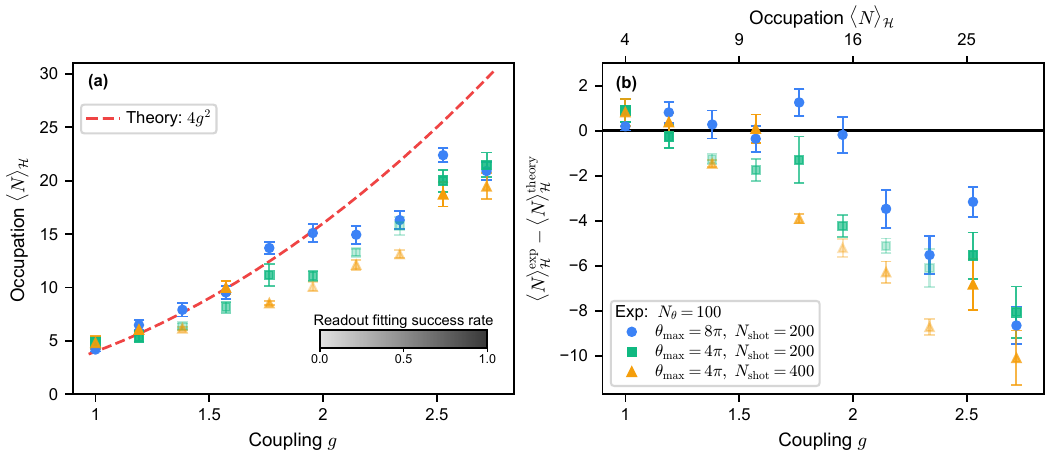}
  \caption{Coupling scan of the static Yukawa evolution $e^{-i \mathcal{H}_s t}$ at fixed $t = \pi$ and flow parameter $s=0$. 
  The initial state has exactly one fermion and zero bosons.
  (a) Occupation $\langle N\rangle_{\mathcal{H}}$ versus coupling $g$, with the coherent-state prediction $4g^2$ (dashed) shown in \eqref{eqn:boson-number}.
  (b) Residual $\langle N\rangle_{\mathcal{H}}^{\rm exp}-\langle N\rangle_{\mathcal{H}}^{\rm theory}$ versus $g$. 
  The top axis reads the corresponding $\langle N\rangle_{\mathcal{H}}=4g^2$.
  Markers distinguish the three runs and error bars are the bootstrap standard deviation. The marker opacity encodes the fitting success rate during the bootstrap. }
  \label{fig:scan-g}
\end{figure*}

We next characterize the readout range by scanning the coupling strength $g$.
We set the time to the peak-amplitude point, $t = \pi$, and the flow parameter to $s=0$, which simulates a bare system.
The result is shown in \cref{fig:scan-g}.

In addition, we tested different maximum angles $\theta_\mathrm{max}$ and total shots per point of the AJC$(\theta)$ for the fit.
All three experiments show a similar trend. The occupation inferred from the reconstructed distribution diverges around 10 to 15 and breaks down completely beyond that.
Although doubling $\theta_\mathrm{max}$ to $8\pi$ improves the fitting success rate significantly, it does not help recover occupations above about 15 quanta to a substantial degree. This indicates an intrinsic fitting difficulty, and the reliable-readout limit we observe is robust.

This behavior is expected, because the reconstruction relies on distinguishing adjacent occupation levels, whose sideband frequencies (\cref{eq:oscillate}) are proportional to $\sqrt{n+1}$ and draw closer together as the occupation grows~\cite{meekhof1996nonclassical, leibfried1996determination, than2026observation}.

The other two scans, over $t$ and $s$, also use $\theta_\mathrm{max}=8\pi$, where the fitting success rate is consistently above 90\%. 
For those two scans we also drop the opacity colormap and display only the points that fit successfully.

\subsection{Scanning flow parameter $s$}
In this experiment we examine the readout of the occupation $\langle N\rangle_{\mathcal{H}}$ for large coupling $g$ by scaling the flow parameter $s$.
We chose $g=3$, which lies beyond the reliable readout range at $s=0$ (\cref{fig:scan-g}).
We then fix $t = \pi$ and scan for flow parameter $s$.
The result, shown in \cref{fig:flow}, demonstrates that increasing the flow parameter scales $\langle N\rangle_{\mathcal{H}}$ down into the readable range. The measured occupation tracks $36\,e^{-2s}$ for $s\gtrsim 0.4$, whereas at $s=0$ and $s=0.2$ the predicted occupation ($\approx 36$ and $24$) lies above the readout range and is undershot.

\section{Discussion}\label{sec:discussion}

\begin{figure}[t]
  \centering
  \includegraphics[width=0.48
\textwidth]{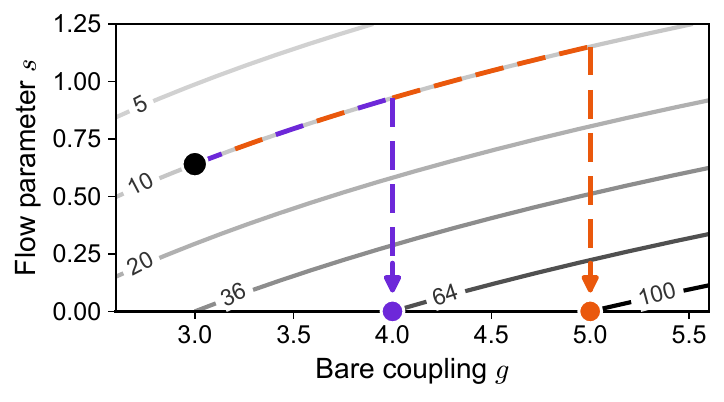}
  \caption{Reinterpreting a single measured occupation $\langle N\rangle_{\mathcal{H}} = 4g^2 e^{-2s}$.
   Gray contours are lines of
  constant occupation $\langle N\rangle_{\mathcal{H}}$ over bare coupling $g$ and
  flow parameter $s$, labeled by value. An example reading $\langle N\rangle_{\mathcal{H}} = 10$ at $g = 3$
  (black dot) is reassigned to a stronger bare coupling by sliding along the contour
  (dashed) to $g = 4$ or $g = 5$, then dropping at fixed $g$ to $s = 0$, recovering the bare
  occupation $\langle N\rangle_H = 4 g^2 = 64$ (purple) and $100$ (orange). }
  \label{fig:reinterpret}
\end{figure}

The occupation ceiling we reach is a property of the readout method, not of the occupation itself.
We reconstruct the occupation from the sideband oscillations of a probe qubit, and the frequencies that label neighboring Fock states crowd together as the occupation grows, so a broad distribution becomes progressively harder to resolve~\cite{meekhof1996nonclassical, leibfried2003quantum}.

The same method, and the same kind of limit, appear in the most directly comparable trapped-ion realization of these dynamics, which resolved occupations up to about eight~\cite{than2026observation}.
Other qubit-boson platforms read the boson by other means, from the dispersive number-resolved readout of circuit QED~\cite{schuster2007resolving, blais2021cqed} to the motional tomography of neutral atoms~\cite{brown2022tomography}, none of them resolving an unknown occupation much beyond ten.
Their limits arise from different, platform-specific mechanisms rather than a shared fundamental bound.
Ours is set by how long the motion stays coherent while the sideband frequencies are resolved, so a longer motional coherence time or a faster sideband drive would raise the ceiling we observe.

The renormalization group flow used here is complementary to such improvements and available on present hardware.
In this experiment the only readout is the occupation $\langle N\rangle_{\mathcal{H}}$, which depends only on the $\beta$ used in the displacement $D(\beta)$.
On the other hand, in the static Yukawa model, $\langle N\rangle_{\mathcal{H}}$ depends only on the renormalized coupling
constant $g_s$, which in turn depends on $s$ and the bare coupling $g$.
Hence we have a single experimental parameter $\beta$ but two theoretical parameters, $g$ and $s$, which enter it only through $g_s$.

Our reconstruction of the occupation distribution is reliable only up to roughly $\langle N\rangle_{\mathcal{H}} = 10$ to $15$.
Increasing $g$ pushes $\langle N\rangle_{\mathcal{H}}$ outside this range, where the measured values become inaccurate.
However, we can reinterpret a well-in-range reading, $\langle N\rangle_{\mathcal{H}} < 10$, as the occupation at a stronger bare coupling.
A larger $g$ with a compensating larger $s$ leaves the renormalized coupling $g_s = g\,e^{-s}$ of \cref{eqn:gs} unchanged, so the same reading applies at that coupling.
The bare occupation there follows from the recovery relation,~\cref{eqn:recover}.

This reinterpretation, shown in \cref{fig:reinterpret}, is a straightforward renormalization group extrapolation.
Such extrapolations are used in practice, for example when computing parton distribution functions (PDFs) from lattice gauge theory~\cite{ji2013parton, ji2021large}.

The recovery, though simple, is not free.
The factor $e^{-2s}$ that scales the occupation into the readout range is undone by $e^{2s}$ on the way back, so the uncertainty of the recovered value grows by the same factor.
At large $s$ the motional heating floor limits how far the occupation can be scaled before the signal is lost~\cite{turchette2000heating}.

Generalizing the construction faces two obstacles.
The first is compiling the renormalized evolution.
Here that evolution reduces to a displacement.
A more demanding case arises when the operator basis is not closed under the flow, because then the flow transformation itself must be implemented.
The flow can in this case be solved perturbatively for small $s$, as in~\cite{serafin2025secondorder, gustin2025renormalized, serafin2026secondorder}.
The second is recovering the bare occupation.
Here that recovery is multiplication by a single closed-form factor.
A more demanding case arises when more than one quantity flows with $s$, because a single reading determines one running coupling but not several independent ones.
We leave this and other generalizations to
future work.

The static Yukawa model is exactly solvable, so this demonstration is a controlled benchmark rather than a computation beyond classical reach.
The occupations recovered at $g=4$ and $g=5$ in \cref{fig:reinterpret} are reinterpretations of an in-range measurement along the known flow, not independent readings at those couplings.
This approach is valuable for models whose renormalized evolution is classically hard yet still implementable on the processor, and whose recovery of the bare quantities stays simple to compute.

\begin{acknowledgments} 
This material is supported by the US Department of Energy, Office of Science, Office of Advanced Scientific Computing Research under its Quantum Testbed Program. Sandia National Laboratories is a multimission laboratory managed and operated by National Technology \& Engineering Solutions of Sandia, LLC, a wholly owned subsidiary of Honeywell International Inc., for the US Department of Energy's National Nuclear Security Administration under contract DE-NA0003525. This paper describes objective technical results and analysis. Any subjective views or opinions that might be expressed in the paper do not necessarily represent the views of the US Department of Energy or the United States Government.

The fitting, bootstrap, and figure code was written by Claude Opus 4.8, Opus 5 and Opus 5.5 (Anthropic) under our direction, and we verified it by reviewing its results. We also used Claude to check our derivations and polish the manuscript.
\end{acknowledgments}

\section*{Data availability}
The data that support the findings of this article are not publicly available. The data are available from the authors upon reasonable request.

\section*{Conflict of interest}
P.J.L. is Chief Scientific Officer at QMatter, Inc.

\clearpage

\appendix
 
\newpage

\section{Hardware and calibration}\label{app:hardware}

The bosonic mode of the static Yukawa model is realized physically by a shared motional (phonon) mode of the trapped-ion chain. To reconstruct the boson's state, we use the anti-Jaynes-Cummings (AJC) interaction~\cite{jaynes1963comparison, leibfried2003quantum} to map the population of individual Fock states onto distinct, measurable oscillation frequencies of a probe qubit.
Specifically, the AJC Hamiltonian, 
\begin{align}
H_{\rm AJC}(\phi)=\frac{\Omega}{2}\left(e^{-i\phi}\,\sigma_{+}a^{\dagger}+e^{i\phi}\,\sigma_{-} a\right),
\end{align}
with probe raising and lowering operators $\sigma_{\pm}$, sideband Rabi rate $\Omega$, and drive phase $\phi$,
drives transitions between the states $|0,n\rangle$ and $|1,n+1\rangle$
\begin{align}
\begin{array}{ccc}
\text{Higher Energy} & & \\
\vdots & \vdots & \vdots \\
|0, 3\rangle & \xleftrightarrow{\quad \Omega_3 \propto \sqrt{4} \quad} & |1, 4\rangle \\
|0, 2\rangle & \xleftrightarrow{\quad \Omega_2 \propto \sqrt{3} \quad} & |1, 3\rangle \\
|0, 1\rangle & \xleftrightarrow{\quad \Omega_1 \propto \sqrt{2} \quad} & |1, 2\rangle \\
|0, 0\rangle & \xleftrightarrow{\quad \Omega_0 \propto \sqrt{1} \quad} & |1, 1\rangle \\
  & & |1, 0\rangle \\
\text{Lower Energy} & & \\
\end{array}
\end{align}
For each initial phonon population $n$, the system is restricted to the $\{|0,n\rangle, |1,n+1\rangle\}$ subspace, on which $H_{\rm AJC}$ generates a rotation by $\alpha_n=\theta\sqrt{n+1}$ about the equatorial axis $(\cos\phi,\sin\phi,0)$, with $\theta\equiv\Omega t$ the commanded pulse area. Because the probe begins in $|0,n\rangle$ and the readout is a population, the transferred probability depends only on $\alpha_n$, not on the drive phase $\phi$. We therefore set $\phi=0$, an $X$-rotation $R_x(\alpha_n)$. Applying it to the initial state $|0,n\rangle$ yields
\begin{align}
  |\psi_n\rangle = \cos\left(\frac{\alpha_n}{2}\right) |0,n\rangle - i \sin\left(\frac{\alpha_n}{2}\right) |1,n+1\rangle.
\end{align}

The probability of measuring the qubit in the excited state given that it started with $n$ phonons is the squared norm of the amplitude attached to $|1,n+1\rangle$,
\begin{align}
  P_{1,n} = |\langle 1,n+1 | \psi_n \rangle|^2 = \sin^2\left(\frac{\alpha_n}{2}\right).
\end{align}
Applying the standard trigonometric half-angle identity, we expand this to
\begin{align}
  P_{1,n} = \frac{1}{2} - \frac{1}{2}\cos(\alpha_n) = \frac{1}{2} - \frac{1}{2}\cos(\theta\sqrt{n+1}).
\end{align}
 
In the QSCOUT testbed~\cite{clark2021qscout}, the trap contains a chain of two trapped ions, providing two physical qubits coupled to shared motional modes. In this particular case, we utilize the tilt mode from the lower radial manifold due to its lower heating rate, $\dot{\bar{n}}\sim 3$--$10$~quanta/s. Methods for the qubit-boson gates utilized on QSCOUT are described in~\cite{tortorici2026qubitboson}.
We assign one of the two qubits as the interaction qubit on which the conditional displacements and Hadamard gates are performed, while the other qubit acts as the probe qubit for reading out the Fock occupancy. Specifically, we post-select the $\{|00\rangle,|10\rangle\}$ subspace in which the interaction qubit remained unchanged, which maps directly onto the logical $\{|0\rangle,|1\rangle\}$ of the probe qubit.
The quantity of interest is $P_{|1\rangle}(\theta)$, the probability that the probe qubit is found in $|1\rangle$.
On hardware we estimate it from the post-selected counts as the excitation fraction
\begin{align} 
  P_{\rm exc}(\theta)=\frac{P_1(\theta)}{P_0(\theta)+P_1(\theta)},
  \label{eq:postselect}
\end{align}
with $P_0(\theta)$ and $P_1(\theta)$ the observed probabilities of $|0\rangle$ and $|1\rangle$.
Post-selection discards only the shots that leak out of the $\{|0\rangle,|1\rangle\}$ readout subspace, so $P_{\rm exc}$ is the hardware estimate of $P_{|1\rangle}$ and is specific to the readout, not to the model.

$P_{|1\rangle}(\theta)$ is a sum of independent Rabi oscillations, one per manifold $\{|0,n\rangle,|1,n+1\rangle\}$ and weighted by its Fock population $p_n$, because the manifolds do not mix~\cite{meekhof1996nonclassical, leibfried1996determination}.
To resolve these overlapping frequencies, we sweep a sequence of $N_\theta$ rotation angles $\{\theta_i\}_{i=1}^{N_\theta}$.

At each scanned angle $\theta_i$, the sequence is repeated $N_\mathrm{shot}$ times to accumulate statistics, for a total of $N_\theta\times N_\mathrm{shot}$ shots across the scan.

\subsection{Readout fit}\label{app:readout}
To extract the underlying bosonic distribution $p_n$, we fit the post-selected data $P_{\rm exc}(\theta_i)$ to the sideband model for $P_{|1\rangle}(\theta)$ by non-linear least squares,
\begin{align}
  P_{|1\rangle}(\theta)=\frac{1}{2}-\frac{1}{2}\,e^{-\Gamma\theta}\sum_{n=0}^{n_{\rm cutoff}-1}p_n\cos\!\big(\eta\sqrt{n+1}\,\theta\big).
  \label{eq:sideband-app}
\end{align}
The angle-calibration factor $\eta$, which rescales $\theta$ to account for systematic deviations from the programmed drive, is restricted to $\eta\in[0.85,1.05]$.
The decoherence rate carries only the physical requirement $\Gamma\ge0$, and each fit is initialized at $\Gamma_0=0.1$. A substantially larger starting value can trap the strongly coupled scans in a spurious solution with $\langle N\rangle_{\mathcal{H}}\!\approx\!0$.
Each Fock population is bounded, $p_n\in[0,1]$, while the sum is left unconstrained, since the bounded Trust Region Reflective solver admits box constraints but not the equality $\sum_n p_n=1$. The unconstrained sum also absorbs the reduced measurement contrast.

The truncation $n_{\rm cutoff}$ is not fixed in advance but chosen by an exhaustive grid search. A scan of $N_\theta$ angles constrains at most $N_\theta$ independent parameters, so with the two hardware parameters $\eta$ and $\Gamma$ also free, $n_{\rm cutoff}$ cannot exceed $N_\theta-2$. Fitting near that saturation, however, lets spurious high-$n$ populations absorb measurement noise and degrades the fit. We therefore select $n_{\rm cutoff}$ by minimizing the reduced chi-square
\begin{align}
  \chi^2_\nu=\frac{\mathrm{RSS}}{N_\theta-n_{\rm cutoff}-2},
  \label{eq:redchi}
\end{align}
where $\mathrm{RSS}$ is the unweighted residual sum of squares (uniform weights, not variance-weighted) and the denominator $N_\theta-n_{\rm cutoff}-2$ is the residual number of degrees of freedom, the $n_{\rm cutoff}$ populations together with $\eta$ and $\Gamma$ making $n_{\rm cutoff}+2$ fitted parameters.
Penalizing by the remaining degrees of freedom suppresses populations that do not reduce $\mathrm{RSS}$. Because the denominator must stay positive, the grid runs over $n_{\rm cutoff}\in[1,\,N_\theta-3]$ (at least the $n{=}0$ population), no maximum cutoff is imposed, and the minimizer is interior for every measured point.

\subsection{Acceptance criteria}\label{app:accept}
A completed fit is accepted as valid only when two conditions hold: (i) the least-squares optimizer converges, and (ii) the calibration factor $\eta$ does not rest on either of its bounds.
Resamples that fail either test are excluded from the bootstrap distribution of \cref{app:error} and lower the per-point success rate.

\section{Error analysis}\label{app:error}
We attach an error bar to each measured point by bootstrapping the recorded shots.
The fit of \cref{app:readout} returns one value of $\langle N\rangle_{\mathcal{H}}$ per scan with no uncertainty of its own, so we take the error bar from the spread of the shots rather than from the fit covariance.
We resample each angle of the sideband scan by drawing a fresh outcome count from its full recorded $4$-outcome distribution at the original shot number, post-select each draw as in \cref{eq:postselect}, and refit the scan with $n_{\rm cutoff}$ held fixed at the value chosen by a single fit to the original data.

Only resamples that pass the acceptance gate of \cref{app:accept} are valid and enter the bootstrap distribution.
Starting from $N_{\mathrm{boot}}=100$ resamples we repeatedly double $N_{\mathrm{boot}}$, reusing the earlier resamples, until both the mean and the standard deviation of the resampled $\langle N\rangle_{\mathcal{H}}$ move by less than $0.05$ quanta on two consecutive doublings, up to a cap of $N_{\mathrm{boot}}=12800$.
We report each point as the mean of its valid resamples, with an error bar of one standard deviation.
The doubling rule monitors the same mean and standard deviation that we report, so both have converged when the rule stops.
This error bar captures the statistical uncertainty from shot noise alone.

The per-point success rate is the fraction of resamples that are valid.
In the coupling scan the success rate falls where the $\theta_{\mathrm{max}}=4\pi$ fits drive the calibration factor $\eta$ to its bound.

We encode the success rate as the marker opacity, and draw solid markers in the time and flow scans.

\section{Exact PRD decomposition for static Yukawa evolution}\label{app:prd}

The renormalized time-evolution $U(t)=e^{-i\mathcal{H}_s t}$ of the static Yukawa model is exactly decomposed into a scalar phase $P$, a phase-space rotation $R$, and a displacement operator $D$.
This decomposition is the Wei-Norman form~\cite{wei1963lie} of the evolution operator, a product of exponentials of oscillator-algebra generators.
We obtain this form by completing the square rather than by integrating the Wei-Norman equations.
To show this, we first derive the time-evolution of the generic driven harmonic oscillator, and then map the variables to $\mathcal{H}_s$.

\subsection{Generic driven harmonic oscillator form}
Consider the generic driven harmonic oscillator Hamiltonian,
\begin{align}
    H = \epsilon + \omega a^\dagger a + \lambda(a^\dagger + a)
\end{align}
where $\omega > 0$ and all parameters ($\epsilon$, $\omega$, $\lambda$) are strictly real numbers.
By completing the square, we rewrite $H$ exactly in terms of a shifted ladder operator:
\begin{align}
    H = \tilde{\epsilon} + \omega \tilde{a}^\dagger \tilde{a}
\end{align}
where we define the real shift amplitude $\alpha$, the shifted operators $\tilde{a}, \tilde{a}^\dagger$, and the shifted ground-state energy $\tilde{\epsilon}$ as:
\begin{align}\alpha &:= -\frac{\lambda}{\omega} \\
    \tilde{a} &:= a - \alpha; \quad \tilde{a}^\dagger := a^\dagger - \alpha\\
     \tilde{\epsilon} &:= \epsilon - \omega \alpha^2.
    \end{align}

\subsection{Mapping to the unshifted frame}
The time-evolution operator for the diagonalized Hamiltonian is 
\begin{align}
    e^{-i H t} &= e^{-i ( \tilde{\epsilon} + \omega \tilde{a}^\dagger \tilde{a})t} \notag \\
&= e^{-i \tilde{\epsilon} t} e^{-i \omega \tilde{a}^\dagger \tilde{a} t}.
\end{align}
To map the shifted frame back to the standard unshifted frame, we use the displacement operator 
\begin{align}
D(\alpha) = e^{\alpha a^\dagger - \alpha^* a},
\end{align} which cleanly executes this shift
\begin{align}
    D(\alpha) a D^\dagger(\alpha) = a - \alpha = \tilde{a}
\end{align}
Note that the inverse is 
\begin{align}
    D^\dagger(\alpha) = D(-\alpha).
\end{align}
The operator portion of the time evolution maps back as follows
\begin{align}
    e^{-i \omega \tilde{a}^\dagger \tilde{a} t} 
    &= D(\alpha) e^{-i \omega a^\dagger a t} D^\dagger(\alpha) \notag \\
    &= D(\alpha) R D(-\alpha)
\end{align}
where we define the standard phase-space rotation 
\begin{align}
    R := e^{-i \omega t a^\dagger a}.
\end{align}

\subsection{Decomposition via operator identities}
To isolate $R$ on the left and merge the remaining adjacent displacement operators, we apply two standard displacement-operator identities, as used in the hybrid oscillator-qubit compilations of Refs.~\cite{crane2024, liu2026}:
\begin{align}
    \label{eqn:dr-rd-generic}
    D(\mu) R &= R D(\mu e^{i\omega t}) \\
    \label{eqn:weyl-generic}D(\mu)D(\nu) 
   & = e^{\frac{1}{2}(\mu\nu^* - \mu^*\nu)} D(\mu+\nu)\end{align}
    Applying these sequentially to our expression gives:
\begin{align}e^{-i \omega \tilde{a}^\dagger \tilde{a} t} &= \underbrace{D(\alpha) R}_{\eqref{eqn:dr-rd-generic}} D(-\alpha) \notag \\
&= R \underbrace{D(\alpha e^{i\omega t}) D(-\alpha)}_{\eqref{eqn:weyl-generic}} \notag \\
&= R \left( e^{-i \alpha^2 \sin(\omega t)} D\big(\alpha(e^{i\omega t} - 1)\big) \right)
\end{align}

\subsection{Final PRD form}
Substituting this operator sequence back into the full time-evolution expression, we collect the scalar phases into $P$:
\begin{align}
    e^{-i H t} &= e^{-i \tilde{\epsilon} t} \left[ R \left( e^{-i \alpha^2 \sin(\omega t)} D\big(\alpha(e^{i\omega t} - 1)\big) \right) \right] \notag \\
    &= \underbrace{\exp\left[-i \big(\tilde{\epsilon} t + \alpha^2 \sin(\omega t)\big)\right]}_{=: P} R \underbrace{D\big(\alpha(e^{i\omega t} - 1)\big)}_{=: D} \notag \\
    &= P(\phi_t) R D(\beta_t)\end{align}

Finally, by expanding the phase $\phi_t$ using our earlier definition $\tilde{\epsilon} = \epsilon - \omega \alpha^2$, we arrive at the exact, simplified generic parameters:
\begin{align}
    \phi_t &= \epsilon t - \alpha^2 \big(\omega t - \sin(\omega t)\big) \\
    \beta_t &= \alpha (e^{i\omega t} - 1).
\end{align}

 \subsection{Exact decomposition of $\mathcal{H}_s$}
 With the generic shifted harmonic oscillator decomposition established, we can immediately apply it to the static Yukawa Hamiltonian by mapping the generic variables to the physical parameters:
 \begin{align}
    \epsilon &\to \epsilon_s = E_{\mathrm{f}} + g_s^2 \Delta_s \\
    \omega &\to \omega = E_{\mathrm{b}} \\
    \lambda &\to \lambda_s = g_s E_{\mathrm{I}}.
\end{align}
This directly yields the static ground-state displacement amplitude,\begin{align}\alpha_s = -\frac{\lambda_s}{\omega} = -\frac{g_s E_{\mathrm{I}}}{E_{\mathrm{b}}},\end{align}and the shifted ground-state energy,\begin{align}\tilde{\epsilon}_s  
    &= \epsilon_s - \omega \alpha_s^2 \notag \\
&= E_{\mathrm{f}} + g_s^2 \Delta_s - \frac{g_s^2 E_{\mathrm{I}}^2}{E_{\mathrm{b}}}.
\end{align}
Substituting these specific parameters into our generic formula $e^{-i \mathcal{H}_s t} = P R D$, the time-evolution operator for the static Yukawa model is exactly decomposed as:\begin{align}e^{-i \mathcal{H}_s t} = P(\phi_t) R(\theta_t) D(\beta_t)\end{align}
By expanding the terms with the physical parameters defined above, we  recover the time-dependent phase $\phi_t$, rotation angle $\theta_t$, and displacement amplitude $\beta_t$ shown in the main text:
\begin{align}\phi_t &= \epsilon_s t - \alpha_s^2 \big(\omega t - \sin(\omega t)\big) \notag \\
    &= (E_{\mathrm{f}} + g_s^2 \Delta_s) t - \frac{g_s^2 E_{\mathrm{I}}^2}{E_{\mathrm{b}}^2} \big(E_{\mathrm{b}} t - \sin(E_{\mathrm{b}} t)\big) \\
    \theta_t &= \omega t = E_{\mathrm{b}} t \\
    \beta_t &= \alpha_s (e^{i\omega t} - 1)  = \frac{g_s E_{\mathrm{I}}}{E_{\mathrm{b}}} \left(1 - e^{i E_{\mathrm{b}} t}\right).\end{align}

\section{Recovery of the bare occupation from the renormalized dynamics: General Identity}
\label{app:general}

We derive the exact algebraic relationship between the bare and renormalized occupations by using explicit unitary rotations, requiring no assumptions about the underlying model. We write the bare unitary evolution of the mode operator by adding and subtracting the renormalized evolution:
\begin{align}
  e^{iHt} a\, e^{-iHt} = e^{i\mathcal{H}_s t} a\, e^{-i\mathcal{H}_s t} + \Xi_s(t),
\end{align}
where
\begin{align}
\Xi_s(t) = e^{iHt} a\, e^{-iHt} - e^{i\mathcal{H}_s t} a\, e^{-i\mathcal{H}_s t}.
\end{align}
Dropping the $s$ subscripts for clarity and
substituting this into the definition of the bare occupation yields
\begin{align}
\langle N\rangle_H &= \langle\mathrm{init}|\, e^{iHt} a^\dagger a\, e^{-iHt} \,|\mathrm{init}\rangle \notag \\
&= \langle\mathrm{init}|\, \Big(e^{i\mathcal{H} t} a^\dagger e^{-i\mathcal{H}  t} + \Xi ^\dagger\Big) \notag \\
& \quad \quad\quad\quad\Big(e^{i\mathcal{H}  t} a\, e^{-i\mathcal{H}  t} + \Xi \Big) \,|\mathrm{init}\rangle \notag \\
&= \underbrace{\langle\mathrm{init}|\, e^{i\mathcal{H}  t} a^\dagger a\, e^{-i\mathcal{H}  t} \,|\mathrm{init}\rangle}_{=\langle N\rangle_{\mathcal{H}}} \notag \\
&\quad + \langle\mathrm{init}|\, e^{i\mathcal{H}  t} a^\dagger e^{-i\mathcal{H}  t} \Xi  \,|\mathrm{init}\rangle \notag \\
&\quad + \langle\mathrm{init}|\, \Xi ^\dagger e^{i\mathcal{H}  t} a\, e^{-i\mathcal{H}  t} \,|\mathrm{init}\rangle \notag \\
&\quad + \langle\mathrm{init}|\, \Xi ^\dagger \Xi  \,|\mathrm{init}\rangle  \notag \\
& = \langle N\rangle_{\mathcal{H}} + 2\,\mathrm{Re}\big\langle\mathrm{init}|\, e^{i\mathcal{H} t} a^\dagger e^{-i\mathcal{H} t} \Xi \,|\mathrm{init}\rangle  \notag \\
&\quad  + \langle\mathrm{init}|\, \Xi^\dagger \Xi \,|\mathrm{init}\rangle \label{eqn:recovery}
\end{align}

\section{Recovery of the bare occupation from the renormalized dynamics: Static Yukawa}
\label{app:yukawa}
For the proof of concept, we can analytically evaluate the term in \cref{eqn:recovery}.
Notice all operators to be evaluated share the same similarity transformation form.
As a result, we explicitly show the result of one example, and the rest follows.

\subsection{Compute $ e^{iHt} a\, e^{-iHt}$}
We take advantage of the  $e^{-iHt} = P(\phi) R(\theta) D(\beta)$ decomposition and use the following identities,
\begin{align}
  R(\theta)^\dagger a R(\theta) &= a e^{-i\omega t}\label{eqn:rar} \\
  D^\dagger(\beta) a  D(\beta) &= a + \beta \label{eqn:dad}.
\end{align}
So,

\begin{align}e^{iHt} a e^{-iHt} &= (D^\dagger R^\dagger \underbrace{P^\dagger) a (P}_{[a,P]=0} R D) \notag \\
  &= D^\dagger \underbrace{R^\dagger a R}_{\eqref{eqn:rar}} D \\
& =  e^{-i\omega t} \underbrace{D^\dagger a D}_{\eqref{eqn:dad}}\\
& =  e^{-i\omega t} (a + \beta)
\end{align}

\subsection{Compute $ e^{i\mathcal{H}t} a\, e^{-i\mathcal{H}t}$ and $ e^{i\mathcal{H}t} a^\dagger\, e^{-i\mathcal{H}t}$}

Recall from \cref{eqn:mathcal_H} that $\mathcal{H}_s$ takes exactly the same solution form as $H$ with just different coefficients $\epsilon \rightarrow \epsilon_s$, $\lambda \rightarrow \lambda_s$,
\begin{align}
  \mathcal{H}_s = \epsilon_s + \omega a^\dagger a + \lambda_s(a^\dagger + a).
\end{align}
Following the same procedure, we only need to update $\beta \rightarrow \beta_{s}$ where
\begin{align}
\alpha_s = -\frac{\lambda_s}{\omega}, \quad \beta_{s} = \alpha_s(e^{i\omega t} - 1).
\end{align}
So
\begin{align}
e^{i\mathcal{H}_s t} a e^{-i\mathcal{H}_s t} &= e^{-i\omega t}(a + \beta_{s}) \label{eqn:similarity-imathcalHt}\\
e^{i\mathcal{H}_s t} a^\dagger e^{-i\mathcal{H}_s t} &= \left(e^{i\mathcal{H}_s t} a e^{-i\mathcal{H}_s t}\right)^\dagger \notag \\
&= e^{i\omega t}(a^\dagger + \beta_{s}^*).
\end{align}

\subsection{Compute $\langle \Xi^\dagger \Xi\rangle$}

The operator $\Xi$ is the difference between the bare and renormalized Heisenberg evolutions of $a$,
\begin{align}
   \Xi &=  e^{iHt} a\, e^{-iHt} - e^{i\mathcal{H}_s t} a\, e^{-i\mathcal{H}_s t}\notag \\
   & = e^{-i\omega t}(a + \beta) - e^{-i\omega t}(a + \beta_{s})   \notag \\ 
   & =  e^{-i\omega t}(\beta - \beta_{s}) 
\end{align}
This gives
\begin{align}
    \Xi^\dagger & = e^{i\omega t}(\beta^* - \beta_{s}^*). \label{eqn:xi}
\end{align}

To compute $\Xi^\dagger \Xi$,
\begin{align}
\Xi^\dagger \Xi &= |e^{-i\omega t}(\beta - \beta_{s})|^2 \notag \\
&= |\beta - \beta_{s}|^2 \notag \\
&= |\alpha(e^{i\omega t}-1) - \alpha_s(e^{i\omega t}-1)|^2 \notag \\
&= (\alpha - \alpha_s)^2 |e^{i\omega t}-1|^2\\
& = (\alpha - \alpha_s)^2  4\sin^2\left( \frac{\omega t}{2}\right).
\end{align}

Since $\Xi^\dagger \Xi$ is purely a scalar, its vacuum expectation value is simply the scalar itself multiplied by the normalized vacuum inner product $\langle 0|0\rangle = 1$:
\begin{align}
\langle \Xi^\dagger \Xi\rangle &= (\alpha - \alpha_s)^2  4\sin^2\left( \frac{\omega t}{2}\right).
\end{align}
 
\subsection{Compute $\langle N\rangle_{\mathcal{H}}$}
\label{eqn:n2-mathcal-complete}

We compute the time-evolved number operator $N_{\mathcal{H}}$ by substituting the compact similarity transformations derived above:
\begin{align}
  N_{\mathcal{H}} &= (e^{i\mathcal{H}_s t} a^\dagger\, e^{-i\mathcal{H}_s t}) (e^{i\mathcal{H}_s t} a\, e^{-i\mathcal{H}_s t}) \notag \\
  &= \left(e^{i\omega t}(a^\dagger + \beta_{s}^*)\right) \left(e^{-i\omega t}(a + \beta_{s})\right)\\
  & = (a^\dagger + \beta_{s}^*)(a + \beta_{s})\\
  & = a^\dagger a + \beta_{s}^* a + \beta_{s} a^\dagger + |\beta_{s}|^2
\end{align}

When evaluating the vacuum expectation value, any term acting on the vacuum state with an annihilation operator $a$, or acted upon by a creation operator $a^\dagger$ from the left, identically vanishes ($a|0\rangle = 0$ and $\langle 0|a^\dagger = 0$). Only the purely scalar magnitude survives

\begin{align}
  \langle N\rangle_{\mathcal{H}} &= \langle 0|N_{\mathcal{H}}|0\rangle \notag \\
  &= \underbrace{\langle 0|a^\dagger a|0\rangle}_{=0} + \underbrace{\langle 0|a|0\rangle}_{=0} \beta_{s}^* + \underbrace{\langle 0|a^\dagger|0\rangle}_{=0} \beta_{s} + \underbrace{\langle 0|0\rangle}_{=1} |\beta_{s}|^2 \notag \\
  &= |\beta_{s}|^2\\
  & = \alpha_s^2 |e^{i\omega t}-1|^2\\
  & = 4\alpha_s^2 \sin^2\left( \frac{\omega t}{2} \right)
  \label{eqn:n2-mathcal}
\end{align}
Plug in the coefficients in static Yukawa, we get
\begin{align}
  \langle N\rangle_{\mathcal{H}} = 4g_s^2  \frac{E_{\mathrm{I}}^2}{E_{\mathrm{b}}^2} \sin^2\left( \frac{E_{\mathrm{b}} t}{2} \right).
\end{align}

\subsection{Compute $\mathrm{Re}\big\langle\mathrm{init}|\, e^{i\mathcal{H} t} a^\dagger e^{-i\mathcal{H} t} \Xi \,|\mathrm{init}\rangle $ }
 
Using the compact forms derived earlier, we have
\begin{align}
  \left(e^{i\mathcal{H} t} a^\dagger e^{-i\mathcal{H} t}\right) \Xi &= \left( e^{i\omega t}(a^\dagger + \beta_s^*) \right) \left( e^{-i\omega t}(\beta - \beta_s) \right) \notag \\
  &= (a^\dagger + \beta_s^*)(\beta - \beta_s) \notag \\
  &= a^\dagger(\beta - \beta_s) + \beta_s^*(\beta - \beta_s).
\end{align}

Taking the vacuum expectation value:
\begin{align}
  \langle e^{i\mathcal{H} t} a^\dagger e^{-i\mathcal{H} t} \Xi \rangle &= \underbrace{\langle 0| a^\dagger|0\rangle}_{=0}(\beta - \beta_s) + \underbrace{\langle 0|0\rangle}_{=1} \beta_s^*(\beta - \beta_s) \notag \\
  &= \beta_s^*(\beta - \beta_s)\\
  &=\left( \alpha_s(e^{-i\omega t}-1) \right) \left( (\alpha - \alpha_s)(e^{i\omega t}-1) \right)\\
  & = \alpha_s(\alpha - \alpha_s)|e^{i\omega t}-1|^2\\
  & = 4\alpha_s(\alpha - \alpha_s)  \sin^2\left( \frac{\omega t}{2} \right)
\end{align}

\subsection{Compute $\langle N\rangle_H$ from $\langle N\rangle_\mathcal{H}$}
Combined together, 

\begin{align}
  \langle N\rangle_H &=  \langle N\rangle_{\mathcal{H}} + 2\,\mathrm{Re}\big\langle   e^{i\mathcal{H} t} a^\dagger e^{-i\mathcal{H} t} \Xi \rangle  + \langle  \Xi^\dagger \Xi \rangle\notag \\
& =  4\alpha_s^2 \sin^2\left( \frac{\omega t}{2} \right) + 2 \cdot 4\alpha_s(\alpha - \alpha_s)  \sin^2\left( \frac{\omega t}{2} \right) \notag  \\
&\quad +4(\alpha - \alpha_s)^2  \sin^2\left( \frac{\omega t}{2}\right)\\
& = 4\left( \alpha_s^2+ 2\alpha_s(\alpha - \alpha_s) + (\alpha - \alpha_s)^2\right)\sin^2\left( \frac{\omega t}{2} \right)\\
& = 4\alpha^2 \sin^2\left( \frac{\omega t}{2} \right)
\end{align}

The result $\langle N\rangle_H $ can be directly expressed using $ \langle N\rangle_\mathcal{H}$ as  
\begin{align}
   \langle N\rangle_H& = \left(\frac{\alpha}{\alpha_s}\right)^2  \langle N\rangle_\mathcal{H}\\
   & = \left(\frac{g}{g_s}\right)^2  \langle N\rangle_\mathcal{H},
\end{align}
which reproduces \cref{eqn:recover}.
 
\bibliography{refs}

\end{document}